# Symmetry-dictated switching of antiferromagnetic magnon transport in 2D multiferroics

Yibo Liu, Jiale Wang, Jiexiang Wang, Ying Dai[*], Baibiao Huang, Xinru Li, and Yandong Ma[*]

School of Physics, State Key Laboratory of Crystal Materials, Shandong University, Shandanan Str. 27, Jinan 250100, China

*Corresponding author: daiy60@sdu.edu.cn (Y.D.); yandong.ma@sdu.edu.cn (Y.M.)

**Abstract**

While antiferromagnetic magnons in two-dimensional (2D) materials hold immense promise for high-frequency spintronics, achieving their efficient active control remains a critical challenge. Here, we propose a universal mechanism for the nonvolatile ferroelectric (FE) switching of antiferromagnetic magnon transport in 2D multiferroic lattices. Our mechanism relies on coupling the magnon geometric phase to the FE-induced sublattice asymmetry in exchange and Dzyaloshinskii-Moriya interactions. This explicitly breaks the exact compensation of opposite-chirality magnons inherent to collinear antiferromagnets, lifting their spin degeneracy and inducing a highly tunable net Berry curvature. Crucially, reversing the FE polarization deterministically swaps these magnetic asymmetries, which completely inverts the net magnon Berry curvature and the resulting anomalous thermal Hall conductivity. Using first-principles and linear spin-wave theory, we rigorously validate this geometric-phase-driven mechanism in single-layer $CuCr_2Se_4$. Our findings establish a robust paradigm for coupling multiferroicity with the magnon geometric phase, paving the way for nonvolatile and electrically switchable antiferromagnetic magnonics.

**Introduction**

Magnons, the quantized collective excitations of ordered spin textures, offer a transformative, Joule-heating-free paradigm for post-Moore information processing [1-9]. While ferromagnetic systems have traditionally dominated this landscape, antiferromagnetic (AFM) magnonics is rapidly emerging as a superior frontier, distinguished by terahertz-frequency dynamics, immunity to external magnetic perturbations, and the absence of stray fields [10-16]. However, achieving the effective and active manipulation of AFM magnon properties remains a formidable challenge hindering practical applications. Conventional manipulation strategies rely on current-induced spin-transfer or spin-orbit torques; yet, these charge-based approaches inevitably reintroduce severe Ohmic dissipation, thereby completely negating the fundamental low-power premise of magnonics [17-21].

To overcome this dissipation bottleneck, achieving purely electrical manipulation—specifically through nonvolatile voltage gating rather than current injection—stands as a critical frontier in magnonics [22-28]. Two-dimensional (2D) multiferroics serve as a unique platform for this all-electric control, where the coexistence of ferroelectricity and magnetism allows polarization-driven structural transitions to fundamentally reconfigure the underlying magnetic interactions [29-35]. Consequently, ferroelectric (FE) switching provides a deterministic knob to actively invert and modulate the magnon transport behaviors. Nevertheless, translating this concept into efficient electrical switching hinges on establishing a robust microscopic coupling mechanism. Furthermore, while numerous 2D multiferroics have been predicted [29-35], a universal mechanism for the nonvolatile electrical reversal of the intrinsic geometric phase of AFM magnons remains an outstanding challenge.

In this letter, we demonstrate a symmetry-based approach to manipulate AFM magnon transport without relying on dissipative currents. We theoretically demonstrate that the inherent structural distortion associated with a 2D FE phase naturally lifts the spatial equivalence between magnetic sublattices. This inversion-symmetry breaking profoundly alters the local magnetic environments, inducing a strong imbalance in both the intralayer Heisenberg couplings and the out-of-plane Dzyaloshinskii-Moriya interaction (DMI). Because of this ferroelectrically driven disparity, the perfect cancellation of transverse magnon currents—a fundamental restriction in conventional collinear antiferromagnets—is elegantly circumvented, yielding a finite and highly sensitive net Berry curvature. By electrically flipping the spontaneous polarization, the hierarchy of these sublattice-dependent interactions is perfectly interchanged, thereby reversing the macroscopic anomalous thermal Hall signal on demand. We substantiate this magnetoelectric concept through combined first-principles calculations and linear spin-wave theory, utilizing single-layer (SL) CuCr$_2$Se$_4$ as a representative model. These explored phenomenon and insight provide a concrete physical basis for designing nonvolatile gate-tunable magnonic devices.

**Results and Discussion**

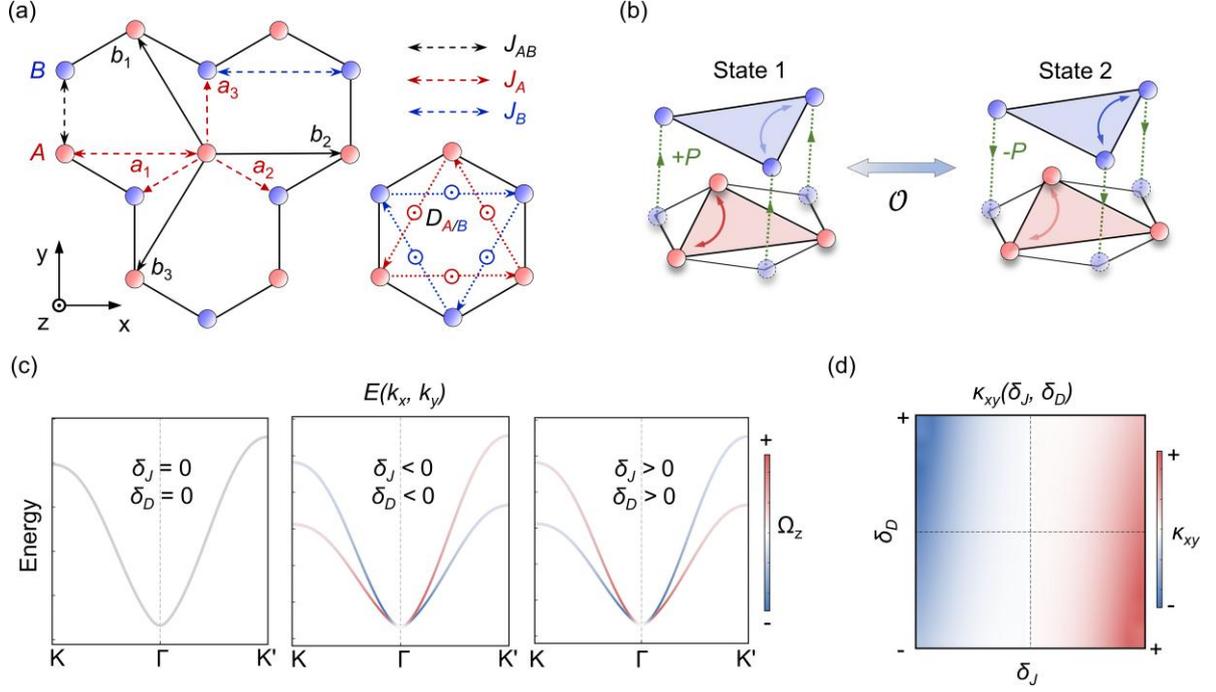

**Figure 1.** (a) Effective spin model on a 2D multiferroic buckled lattice with Néel AFM order. Left: Intralayer and interlayer Heisenberg exchange interaction pathways. Right: Spatial distribution of the out-of-plane DMI vectors. (b) Schematic of the FE switching. Reversing the spontaneous polarization (State 1 ↔ State 2) geometrically swaps the structural environments of the two sublattices. (c) Evolution of magnon band dispersions and local $\Omega_n(k)$ under varying sublattice asymmetry parameters $\delta_J = |J_A| - |J_B|$ and $\delta_D = |D_A| - |D_B|$ conditions. (d) Phase diagram of the anomalous thermal Hall conductivity $\kappa_{xy}$ as a function of $\delta_J$ and $\delta_D$.

The proposed mechanism is based on symmetry-driven magnon geometric phase in 2D multiferroics. To capture the essential spin physics of the 2D multiferroic system, we employ an effective Heisenberg-type spin Hamiltonian on a buckled honeycomb lattice with Néel AFM order. The magnetic interactions are governed by the Hamiltonian:

$$H= J_{inter} \sum_{\langle i,j \rangle} \mathbf{S}_i \cdot \mathbf{S}_j + J_\alpha \sum_{\langle\langle i,j \rangle\rangle} \mathbf{S}_i \cdot \mathbf{S}_j + D_{\alpha ij} \sum_{\langle\langle i,j \rangle\rangle} v_{ij}(\mathbf{S}_i \times \mathbf{S}_j) + K \sum_i (S_i^z)^2, \tag{1}$$

where $\mathbf{S}_{i(j)}$ is the normalized spin vector. The terms sequentially represent the nearest-neighbor inter-sublattice AFM exchange ($J_{inter} > 0$), the intra-sublattice ferromagnetic exchange [$J_\alpha < 0$ ($\alpha$ = A or B)], the symmetry allowed out-of-plane DMI vector ($D_{\alpha ij}$ represent), and the single-ion anisotropy ($K < 0$) establishing an out-of-plane easy axis. $v_{ij} = 2\sqrt{3} a_i \times a_j = \pm \hat{z}$. $a_i$ and $a_j$ are in-plane vectors connecting sites $i$ and $j$ [Fig. 1(a)]. For simplicity, we set the length of the primitive in-plane vectors to be unity [$b_{i/j}$] = 1.

By mapping the spin operators to bosons via the linear Holstein-Primakoff transformation and performing a paraunitary Bogoliubov transformation, we obtain the magnon dispersion spectrum of the buckled honeycomb lattice (Note S1). To establish the fundamental mechanism, we first consider the spatially symmetric lattice where the structurally equivalent A and B sublattices host identical interaction parameters. As depicted in the left panel of Fig. 1(c), the presence of joint inversion ($P$) and a π rotation around the x axis in the spin space ($c_x$) symmetries enforces a strict two-fold spin degeneracy across the entire Brillouin zone. While the inherent DMI generates an imbalance between $+k$ and $-k$ states (allowing individual chirality channels to acquire a geometric phase), the joint $Pc_x$ symmetry dictates that the Berry curvatures $\Omega_n(k)$ of the degenerate bands are exactly opposite in sign. Consequently, the net Berry curvature rigorously vanishes, suppressing any net anomalous thermal Hall conductivity ($\kappa_{xy}$).

In our multiferroic paradigm, the spontaneous FE polarization breaks this spatial equivalence between the sublattices. This inherent structural distortion manifests mathematically as non-zero asymmetry parameters in both the Heisenberg exchange, $\delta_J = |J_A| - |J_B|$, and the DMI, $\delta_D = |D_A| - |D_B|$. Taking State-1 ($\delta_J < 0$ and $\delta_D < 0$) as an example, this structurally induced disparity explicitly breaks the band reciprocity and completely lifts the spin degeneracy [Fig. 1(c), middle panel].

Calculating the Berry curvature via the gauge-invariant Kubo formula reveals that both non-degenerate magnon branches acquire finite Berry curvatures $\Omega_n$ (Note S2). Crucially, the magnitude distribution of $\Omega_n(k)$ becomes highly asymmetric in momentum space, resonantly concentrated around the Γ point due to the minimal energy gap [Fig. 1(c), middle panel]. Under a longitudinal thermal gradient, this Berry curvature acts as an effective magnetic field in momentum space, driving the transverse deflection of thermally populated magnon wave packets [4,36]. The resultant imbalance elegantly bypasses the macroscopic cancellation of transverse currents, yielding a finite net anomalous thermal Hall effect whose magnitude strictly scales with the degree of sublattice asymmetry ($\delta_J$ and $\delta_D$) [Fig. 1(d)].

The deterministic coupling between multiferroicity and magnonics is consummated upon the reversal of the FE polarization (State 2). Reversing the FE polarization naturally interchanges the spatial environments of the top and bottom sublattices [Fig. 1(b)], effectively enforcing the transformation $\delta_J \to -\delta_J$ and $\delta_D \to -\delta_D$. While the energy dispersion is identical to that of State 1, the geometric phase of the bands is fundamentally altered: the sign of the Berry curvature in momentum space is reversed [Fig. 1(c), left panel]. Consequently, the direction of the magnon thermal transport is inverted [Fig. 1(d)]. This mechanism creates a direct coupling between the FE order and the geometric phase physics of the AFM magnons, making the electrical switching of the thermal Hall effect possible.

Synthesizing the above framework, we formulate three universal design criteria to actualize this nonvolatile electrically switchable magnonics platform. First, the lattice must host a buckled geometry separating the magnetic sublattices along the out-of-plane direction. Second, the system must possess a switchable out-of-plane FE polarization to break the joint symmetry. Third, the two FE states must be connected by a symmetry operation $O$ that enforces the reversal of the thermal Hall conductivity, i.e., $O\kappa_{xy}O^{-1} = -\kappa_{xy}$.

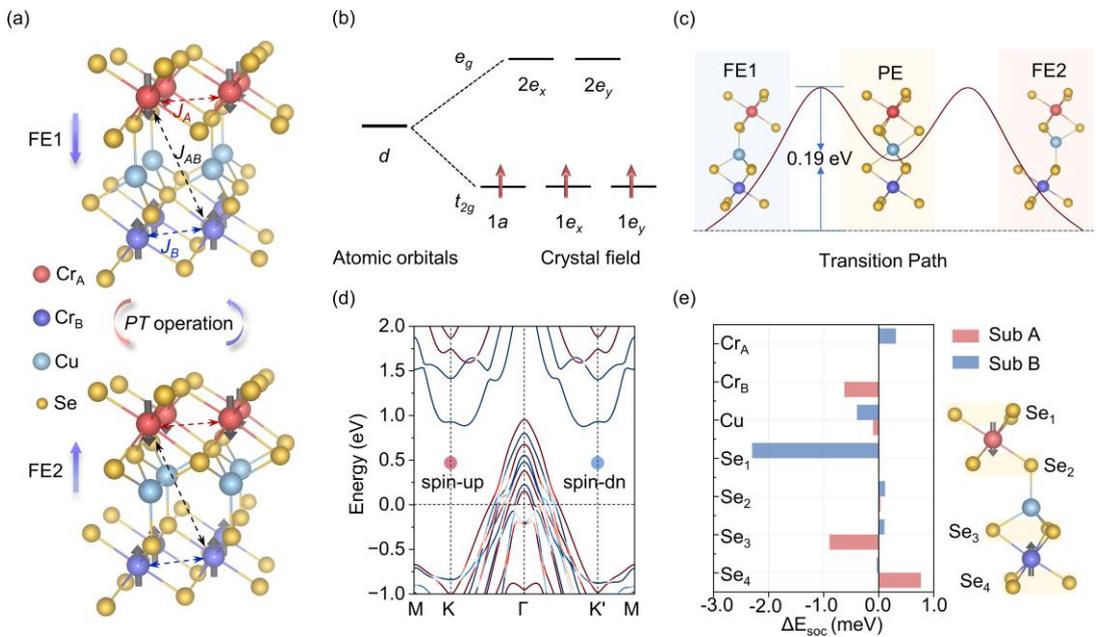

**Figure 2.** (a) Crystal structures of SL $CuCr_2Se_4$ under two degenerate FE states (FE1/2). (b) Diagram

of Cr-*d* orbital occupations. (c) FE transition pathway and associated energy barrier between the FE1 and FE2 states of SL CuCr$_2$Se$_4$. (d) Spin-polarized band dispersions of SL CuCr$_2$Se$_4$. (e) Atomic-projected SOC energy ($\Delta E_{SOC}$) for FE1 state.

Guided by these universal design criteria, we identify the van der Waals multiferroic SL CuCr$_2$Se$_4$ as an ideal physical platform to materialize this nonvolatile magnon switching. SL CuCr$_2$Se$_4$ has been successfully synthesized in recent experiments [37-40]. As depicted in Fig. 2(a), it crystallizes in a hexagonal lattice (space group *P*3*m*1). The lattice constant is optimized to be 3.63 Å, which is in agreement with previous theoretical and experimental findings [41-43]. Structurally, the unit cell comprises two vertically shifted T-phase CrSe$_2$ units intercalated with an off-center Cu atom [Fig. 2(a)]. This intrinsic vertical displacement of the Cu explicitly breaks spatial inversion symmetry, naturally giving rise to two energetically degenerate FE states, FE1 and FE2. To intimately connect with our theoretical model, we assign the Cr atom farther from the Cu dopant in the FE1 state to sublattice A (Cr$_A$) and the closer one to sublattice B (Cr$_B$). The out-of-plane FE polarization is calculated to be P = 2.67 μC cm$^{-2}$ for FE1 state and −2.67 μC cm$^{-2}$ for FE2 state. The transition between these states involves the vertical migration of the Cu atom through the Se plane. As shown in Fig. 2(c), the calculated energy barrier for FE switching is approximately 0.19 eV/f.u., a moderate value that ensures the stability of the FE state while permitting non-volatile switching under practical external electric fields [44,45].

Beyond its FE nature, SL CuCr$_2$Se$_4$ exhibits intrinsic spin polarization, making it a prototypical multiferroic system. The magnetic ground state arises from the Cr atoms, which sit in a distorted octahedral crystal field formed by the Se ligands. As illustrated in the orbital diagram in Fig. 2(b), this crystal field splits the Cr 3*d* orbitals into lower-energy *t$_{2g}$* and higher-energy *e$_g$* states. The three valence electrons of the Cr$^{3+}$ ion half-fill the *t$_{2g}$* sub-shell, resulting in a localized spin moment of S = 3/2 (3 $\mu_B$) according to Hund's rules. Crucially, the out-of-plane FE polarization inherently breaks the spatial equivalence of the magnetic sublattices. In the FE1 state, the distinct coordination environments of Cr atoms induce a strong magnetoelectric coupling, manifested as unequal local magnetic moments of 3.53 $\mu_B$ for Cr$_A$ and 3.61 $\mu_B$ for Cr$_B$. Upon switching to the FE2 state, the spatial position of Cu atom is inverted relative to the Cr layers, causing the magnitudes of the magnetic moments on $Cr_A$ and $Cr_B$ to swap. This deterministic modulation of the magnetic environment directly validates our theoretical premise of polarization-dependent sublattice asymmetry. Furthermore, the spin-polarized band structure presented in Fig. 2(d) reveals that SL $CuCr_2Se_4$ exhibits metallic characteristics, suggesting its potential for spintronic applications involving charge-spin conversion.

**TABLE 1.** Calculated intralayer exchange couplings ($J_A$, $J_B$), interlayer exchange coupling ($J_{inter}$), and out-of-plane DMI strengths ($D_A$, $D_B$) for SL $CuCr_2Se_4$ in the FE1 and FE2 states. All values are given in meV.

|     | $J_A$    | $J_B$    | $J_{inter}$ | $D_A$ | $D_B$ |
|-----|----------|----------|-------------|-------|-------|
| FE1 | −25.400  | −31.810  | 0.971       | 0.229 | 0.588 |
| FE2 | −31.810  | −25.400  | 0.971       | 0.588 | 0.229 |

To quantitatively ground the magnon transport model of SL $CuCr_2Se_4$, we map the total energies from first-principles calculations to the Heisenberg Hamiltonian defined in Eq. (1). Here, we first discuss FE1 state. The extracted magnetic parameters are summarized in Table 1, which provide compelling evidence for the ferroelectrically driven sublattice asymmetry. Specifically, for FE1 state, while the nearest-neighbor intralayer couplings for both sublattices ($J_A$ and $J_B$) are ferromagnetic, the FE-induced structural distortion explicitly breaks their spatial equivalence, rendering their magnitudes distinctly unequal ($\delta_J \neq 0$). Concurrently, the robust AFM interlayer exchange ($J_{inter}$) enforces the requisite Néel-type antiparallel spin alignment between sublattices A and B. Together, these quantitative microscopic parameters unambiguously confirm that the FE1 state of SL $CuCr_2Se_4$ naturally embodies the symmetry-broken "State 1" of our theoretical framework, establishing a realistic platform to actualize the ferroelectrically switchable magnon Berry physics.

Equally crucial to our geometric phase mechanism is the DMI emerging from spin-orbit coupling (SOC). In SL $CuCr_2Se_4$, a mirror plane bisecting the neighboring Cr-Cr bonds dictates the symmetry of the DMI vectors. According to the Moriya's rule [46], the DMI vector can be expressed as $\mathbf{D}_{ij} = d_\parallel(\hat{z} \times \mathbf{u}_{ij}) + d_\perp \hat{z}$, where $\hat{z}$ is the unit vector along z-direction, and $\mathbf{u}_{ij}$ denotes the unit vector pointing from sites i to j (Fig. S1). For the collinear Néel ground state considered here, the magnon excitation spectrum is exclusively governed by the out-of-plane DMI component. Accordingly, all subsequent references to DMI specifically denote the out-of-plane component unless otherwise stated. To quantify the sublattice-resolved DMI strengths, we use a $\sqrt{3} \times 2$ supercell and consider four different noncollinear spin configurations, as shown in Fig. S2. As detailed in Table 1, although the out-of-plane DMI vectors for sublattices A and B share the same polarity, the FE-induced structural distortion provokes a pronounced disparity in their magnitudes. Specifically, for FE1 state, the DMI strength on sublattice B is approximately twice that on sublattice A. This massive inequality ($\delta_D \neq 0$) acts in concert with the Heisenberg exchange asymmetry ($\delta_J$).

To elucidate the microscopic origin of this critical DMI asymmetry, we calculate the atom-resolved contribution to the SOC energy difference, defined as $\Delta E_{SOC} = E_{SOC1} - E_{SOC2}$, where $E_{SOC1}$ and $E_{SOC2}$ represent the energies of the Heisenberg system under different noncollinear spin states (Fig. S2). As illustrated in Fig. 2(e), the vertical displacement of the Cu atom severely alters the local coordination environments, manifesting as competing atomic DMI contributions. For sublattice A in FE1 state, the DMI originates predominantly from $Cr_B$ and $Se_3$ atoms, but suffers strong suppression from Se4. In stark contrast, all adjacent selenium atoms (governed largely by Se1) constructively enhance the DMI on sublattice B. This disparate atomic-level interference elegantly dictates the significantly stronger DMI on sublattice B. Furthermore, the calculated single-ion anisotropy parameter is calculated to be K = 0.238 meV. Given the relatively small magnetocrystalline anisotropy energy in this monolayer system, the spins are susceptible to reorientation; even a modest external magnetic field can polarize and align the moments along the field direction. We subsequently focus on the magnon band characteristics within the vertical magnetic order.

The crowning feature of SL CuCr$_2$Se$_4$ lies in the deterministic electrical control over these microscopic magnetic asymmetries. Upon switching the spontaneous polarization from FE1 to FE2 state, the off-center Cu atom migrates across the Se midplane, completely swapping the local coordination environments of the $Cr_A$ and $Cr_B$ atoms. As explicitly corroborated by our extracted parameters (Table 1), this precise geometric inversion forcefully interchanges the intralayer Heisenberg exchange strengths ($J_A \leftrightarrow J_B$) and the DMI magnitudes ($D_A \leftrightarrow D_B$) between the two sublattices. Crucially, the interactions defining the overarching magnetic ground state—the interlayer AFM coupling ($J_{inter}$) and the single-ion anisotropy ($K$)—remain strictly invariant against this FE switching. This robust structural reconfiguration guarantees a perfect sign reversal of the asymmetry parameters, driving the system exactly from ($\delta_J$, $\delta_D$) to ($-\delta_J$, $-\delta_D$). By physically materializing the theoretical symmetry operation $O$ defined in our generic model, this polarization flip deterministically inverts the highly concentrated magnon Berry curvature in momentum space. Ultimately, this seamless magnetoelectric coupling bypasses the need for dissipative external currents, enabling a purely electrical, nonvolatile, and Joule-heating-free reversal of the intrinsic anomalous thermal Hall effect.

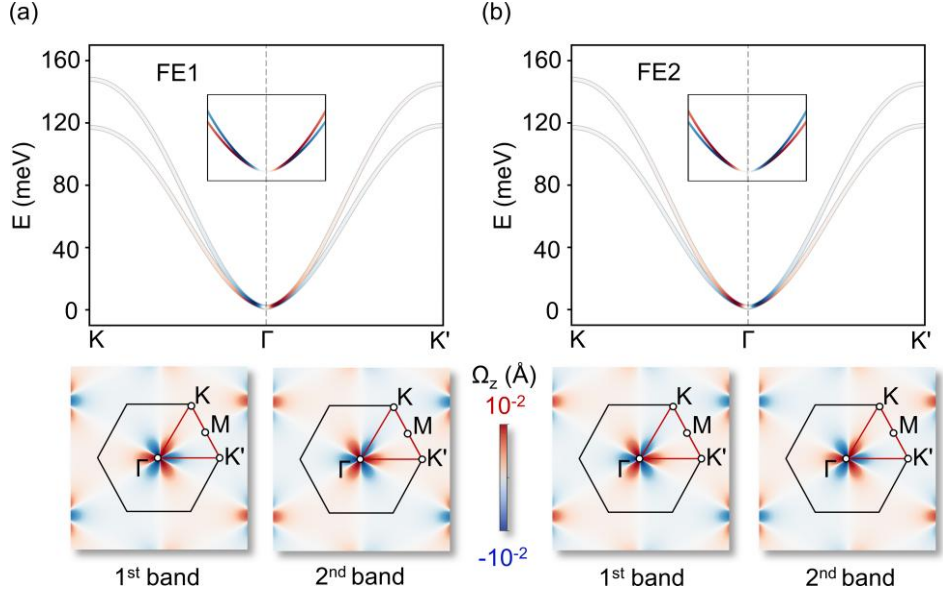

**Figure 3.** Magnon band dispersions and the corresponding Berry curvatures over the 2D Brillouin zone for SL $CuCr_2Se_4$ under (a) FE1 and (b) FE2 states. The detailed band dispersions near the $\Gamma$ point are shown in the insets.

By physically materializing the theoretical symmetry operation $O$ defined in our generic model, this polarization-driven reversal of local asymmetries provides the exact microscopic prerequisite to invert the magnon band character. To quantitatively capture this magnon geometric phase transition, we incorporate these first-principles parameters (Table 1) back into our spin Hamiltonian to evaluate the magnon dispersion of SL $CuCr_2Se_4$.

Figure 3(a) presents the calculated magnon dispersions for the FE1 state. While the spin degeneracy is preserved at the $\Gamma$ point, it is explicitly lifted across the majority of the Brillouin zone, particularly at the K and K' valleys. This momentum-dependent splitting perfectly aligns with our symmetry analysis: the simultaneous presence of the Néel order and the FE-induced structural asymmetry ($\delta_J$ and $\delta_D$) definitively breaks the spatial equivalence that would otherwise protect the degeneracy. Consequently, the separated magnon branches acquire non-zero local Berry curvatures.

As visualized in the inset of Fig. 3(a), these geometric phases are predominantly concentrated in the vicinity of the Γ point. Notably, the upper and lower magnon branches exhibit $Ω(k)$ distributions with opposite signs. The macroscopic transverse response is thus governed by a competitive thermal population mechanism: the incomplete cancellation of these opposing curvatures, weighted by the differential Bose-Einstein occupations of the split energy bands, yields a finite net geometric phase.

Crucially, the FE switching acts as a deterministic control knob for this geometric phase. Upon transitioning to the FE2 state [Fig. 3(b)], the macroscopic energy dispersion remains strictly invariant. However, the momentum-space distribution of the Berry curvature is perfectly inverted [$Ω(k) → –Ω(k)$]. This rigorous inversion is the direct macroscopic manifestation of the swapped microscopic parameters.

To further elucidate the symmetry-driven nature of this effect, we examine the paraelectric (PE) state (Fig. S3). The PE phase restores the centrosymmetric structural lattice. Although the staggered AFM order inherently breaks both *P* and time-reversal (*T*) symmetries individually, the system remains invariant under the joint *PT* operation. This *PT* symmetry imposes a strict constraint on the geometric phase: while *P* transforms the Berry curvature $Ω(k)$ into $Ω(–k)$, the *T* operation further reverses its sign to $–Ω(k)$, collectively necessitating $Ω(k) = 0$ throughout the Brillouin zone. Consequently, the magnon geometric phase vanishes identically in the PE state.

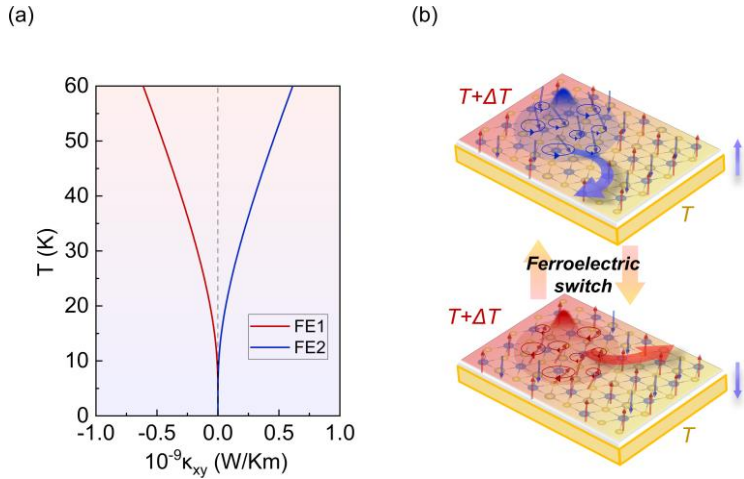

**Figure 4.** (a) The anomalous thermal Hall conductivity $κ_{xy}$ of FE1 and FE2 states for SL $CuCr_2Se_4$. (b) Diagram of the electrically tunable anomalous thermal Hall effect in SL $CuCr_2Se_4$ in the presence of an in-plane thermal gradient field. Red and blue arrows represent the up and down spins.

This symmetry evolution establishes a complete, gate-tunable magnonic architecture: the structural phase transition between the *PT*-symmetric PE state and the *PT*-broken FE state functions as an "ON/OFF" switch for the generation of magnon currents, while the FE1–FE2 transition provides deterministic control over their directionality. To evaluate this ultimate macroscopic consequence, we apply a longitudinal temperature gradient (ΔT) across the 2D lattice. Governed by semiclassical wave-packet dynamics, the non-zero Berry curvature acts as a fictitious magnetic field in momentum space, imparting an anomalous transverse velocity to the propagating magnons, proportional to $\Omega_z(k) \times \nabla T$ [4,36,47].

As schematically illustrated in the inset of Fig. 4(b), for the FE1 state, specific hierarchy of sublattice asymmetries ($\delta_J$ and $\delta_D$) drives a macroscopic net transverse deflection of the thermally populated magnons to the left. This corresponds to a negative thermal Hall conductivity, as plotted in Fig. 4(a). Crucially, upon electrically switching to the FE2 state, the deterministic sign reversal of the asymmetry parameters ($-\delta_J$ and $-\delta_D$) inverts the Berry curvature. Consequently, the transverse magnon current is routed entirely to the right yielding a symmetric and positive $\kappa_{xy}$. Furthermore, the temperature dependence of $\kappa_{xy}$ corroborates its bosonic transport character, i.e., the magnitude is initially negligible at near-zero temperatures, but rapidly increases as higher-energy magnon states with uncompensated Berry curvatures become thermally populated. This macroscopic observable unambiguously confirms that the thermal transport direction is strictly locked to, and fully reversible by, the out-of-plane FE polarization.

Finally, we emphasize that the symmetry-dictated mechanism uncovered here transcends the specific material platform of SL $CuCr_2Se_4$. The essential physical prerequisites—namely a buckled magnetic lattice with broken *P* symmetry, switchable out-of-plane polarization, and finite SOC—form a universal blueprint. This paradigm can be readily extended to a broader family of 2D isostructural multiferroics, including $AgCr_2X_4$ and $NaCr_2X_4$ (X = S, Se) [48,49]. Given the rapid experimental advances in the synthesis of van der Waals magnets, our findings provide a concrete, widely applicable route for realizing nonvolatile, electrically tunable magnonic devices without the crippling constraint of Joule-heating.

**Conclusion**

In summary, we formulate a universal, symmetry-driven mechanism to achieve nonvolatile electrical control over AFM magnon transport in 2D van der Waals multiferroics. By inherently coupling the magnon geometric phase to the ferroelectrically induced sublattice asymmetries ($\delta_J$ and $\delta_D$), we demonstrate that switching the spontaneous out-of-plane polarization effectively executes a spatial inversion, deterministically reversing the net magnon Berry curvature and the ensuing anomalous thermal Hall conductivity. Beyond the concrete first-principles validation in SL $CuCr_2Se_4$, the physical blueprint established here—relying on a buckled magnetic lattice and switchable inversion-symmetry breaking—is highly generalizable to a broad spectrum of isostructural 2D materials. Our study not only uncovers a profound fundamental interplay between structural transitions and magnon geometric phases, but also establishes a robust, Joule-heating-free paradigm for designing next-generation, gate-tunable AFM magnonic architectures.

**Experimental Section**

Our first-principles calculations are performed within the framework of density functional theory (DFT), as implemented in Vienna *ab initio* Simulation Package (VASP) [50,51]. The ionic potential is described by the projected augmented wave (PAW) approach [52]. The exchange-correlation interactions are characterized by the Perdew-Burke-Ernzerhof (PBE) functional of generalized gradient approximation (GGA) [53]. To avoid interactions between adjacent layers, a vacuum layer of more than 15 Å is set along the out-of-plane direction. The cutoff energy and the convergence criterion are set to 520 and $10^{-6}$ eV, respectively. All atoms are fully relaxed until the atomic forces are less than 0.01 eV Å$^{-1}$. A 7 × 7 × 1 Monkhorst–Pack k-point mesh is adopted to sample the 2D Brillouin zone during structural relaxation [54]. SOC is included in the electronic structure calculations. The configuration-dependent total energy difference method is used to extract the exchange coupling strength and DMI strength; this approach has been widely used to determine magnetic parameters [55,56]. For the calculation of exchange interaction parameters, a 1 × 2 supercell with a 15 × 7 × 1 k-point mesh is adopted, while a 7 × 9 × 1 k-point mesh is utilized for the rectangular supercell in the calculation of DMI parameters. The rectangular supercell lattice vectors are defined as $a_s$ = 2a and $b_s$ = √3a along the x and y directions, respectively, where a is the lattice constant of the hexagonal unit cell. The correlation effects of Cr-3*d* electrons are described within the GGA+U method with effective Hubbard U parameter of 5 eV, as adopted in previous work [57].

**Supporting Information**

Figure S1: schematic diagrams of the DMI vectors; Figure S2: noncollinear spin configurations; Figure S3: crystal structure of SL $CuCr_2Se_4$ in PE state; Note S1: Holstein-Primakoff transformation and Bogoliubov transformation; Note S2: Berry curvature and intrinsic anomalous thermal Hall conductivity.

**Conflict of Interest**

The authors declare no conflict of interest.

**Acknowledgements**

This work is supported by the National Natural Science Foundation of China (Nos. 12274261 and 12074217), Shandong Provincial QingChuang Technology Support Plan (No. 2021KJ002), and Taishan Young Scholar Program of Shandong Province.